\documentclass[useAMS,usegraphicx,usenatbib]{mn2e}

\newcommand{\logg}{\ensuremath{\log g}}                    
\newcommand{\Mjup}{\ensuremath{\,{\rm M}_{\rm Jup}}}       
\newcommand{\ms}{\,m\,s$^{-2}$}                            
\newcommand{\mss}{\,m\,s$^{-2}$}                            
\newcommand{\mc}[1]{\multicolumn{2}{c}{#1}}

\title[Direct determination of surface gravities of exoplanets]
      {A method for the direct determination of the surface gravities of transiting extrasolar planets}

\author[Southworth, Wheatley \& Sams]
       {John Southworth\thanks{E-mail: j.k.taylor@warwick.ac.uk}, Peter J Wheatley and Giles Sams \\
        Department of Physics, University of Warwick, Coventry, CV4 7AL, UK}

\begin{document} \maketitle 

\begin{abstract}
We show that the surface gravity of a transiting extrasolar planet can be calculated from only the spectroscopic orbit of its parent star and the analysis of its transit light curve. This does not require additional constraints, such as are often inferred from theoretical stellar models or model atmospheres. The planet's surface gravity can therefore be measured precisely and from only directly observable quantities. We outline the method and apply it to the case of the first known transiting extrasolar planet, HD\,209458\,b. We find a surface gravity of $g_{\rm p} = 9.28 \pm 0.15$\mss, which is an order of magnitude more precise than the best available measurements of its mass, radius and density. This confirms that the planet has a much lower surface gravity that that predicted by published theoretical models of gas giant planets. We apply our method to all fourteen known transiting extrasolar planets and find a significant correlation between surface gravity and orbital period, which is related to the known correlation between mass and period. This correlation may be the underlying effect as surface gravity is a fundamental parameter in the evaporation of planetary atmospheres.
\end{abstract}

\begin{keywords}
stars: planetary systems --- stars: individual: HD\,209458 --- stars: binaries: eclipsing --- stars: binaries: spectroscopic --- methods: data analysis
\end{keywords}


\section{Introduction}

Since the discovery that the star HD\,209458 is eclipsed by a planet in a short-period orbit \citep{Henry+00apj,Charbonneau+00apj} it has become possible to derive the basic astrophysical properties of extrasolar planets and compare these quantities with theoretical predictions \citep[e.g.][]{Baraffe+03aa}. However, the absolute masses, radii and density of the transiting planet cannot be calculated directly from the transit light curve and the velocity variation of the parent star, so extra information is required in order to obtain them. Additional constraints can be found from spectral analysis of the parent star or by imposing a theoretical stellar mass--radius relation \citep{CodySasselov02apj}, but this causes a dependence on theoretical stellar models or model atmospheres. The uncertainties in these constraints dominate the overall errors in mass, radius and density \citep[e.g.][]{Konacki+04apj}, limiting the accuracy with which properties of the star and planet can be measured.

In this work we show that the surface gravity of the planet can be measured directly using only the transit light curve and the radial velocity amplitude of the parent star. No additional information is required and so accurate and precise surface gravity values can be obtained. As theoretical studies often supply predicted values for the surface gravities of planetary objects \citep[e.g.,][]{Baraffe+03aa} we propose that this quantity is very well suited for comparing observation with theory. In addition, the surface gravity is an important parameter in constructing theoretical models of the atmospheres of planets \citep{Marley+99apj,Hubbard+01apj}.

After deriving an equation for surface gravity in terms of directly observed parameters, we illustrate this concept by studying HD\,209458. We then apply it to the other known transiting extradsolar planets using results available in the literature. The planet HD\,209458\,b is known to be oversized for its mass and to be strongly irradiated by the star it orbits. An excellent transit light curve was obtained for HD\,209458 by \citet{Brown+01apj}, who used the HST/STIS spectrograph to obtain high-precision photometry covering several different transit events. Precise radial velocity studies of HD\,209458 are also available \citep{Henry+00apj,Mazeh+00apj,Naef+04aa}.


\begin{table*} \begin{center}
\caption{\label{tab:lcresult} Results of the modelling of the HST/STIS light curve
of HD\,209458. The upper part of the table gives the optimised parameters and the
lower part gives quantities calculated from these parameters. The midpoint of the
transit, $T_{\rm Min\,I}$, is expressed in HJD $-$ 2\,400\,000.}
\begin{tabular}{l r@{\,$\pm$\,}l r@{\,$\pm$\,}l r@{\,$\pm$\,}l r@{\,$\pm$\,}l} \hline
Limb darkening        &    \mc{Linear}      &  \mc{Quadratic}   & \mc{Square-root} & \mc{Adopted parameters}\\
\hline
$r_\star$+$r_{\rm p}$ &  0.12889 & 0.00042  &  0.12771 & 0.00049  &  0.12799 & 0.00050  &      \mc{}        \\
$k$                   &  0.12260 & 0.00011  &  0.12097 & 0.00025  &  0.12051 & 0.00037  &      \mc{}        \\
$i$ (deg.)            & 86.472   & 0.040    & 86.665   & 0.054    & 86.689   & 0.060    & 86.677  & 0.060   \\
$T_0$                 &51659.936716&0.000021&51659.936712&0.000021&51659.936711&0.000021&      \mc{}        \\
$u_\star$             &  0.494   & 0.004    &  0.297   & 0.027    &$-$0.312  & 0.127    &      \mc{}        \\
$v_\star$             &       \mc{ }        &  0.338   & 0.047    &  1.356   & 0.218    &      \mc{}        \\
\hline
$r_\star$             &  0.11481 & 0.00036  &  0.11393 & 0.00042  &  0.11418 & 0.00042  & 0.11405 & 0.00042 \\
$r_{\rm p}$           &  0.01408 & 0.00006  &  0.01378 & 0.00007  &  0.01376 & 0.00008  & 0.01377 & 0.00008 \\
$\chi^2_{\rm \ red}$  &       \mc{1.146}    &     \mc{1.056}      &     \mc{1.054}      &      \mc{}        \\
\hline \end{tabular} \end{center}\end{table*}

\section{Surface gravity measurement}

The fractional radii of the star and the planet in the system are defined as
\begin{equation}
r_\star = \frac{R_\star}{a} \qquad \qquad r_{\rm p} = \frac{R_{\rm p}}{a}
\end{equation}
where $a$ is the orbital semi-major axis, and $R_\star$ and $R_{\rm p}$ are the absolute radii of the star and planet, respectively. $r_\star$ and $r_{\rm p}$ can be directly determined from a transit light curve.

The mass function of a spectroscopic binary is given by \citep[e.g.][]{Hilditch01book}:
\begin{equation}
f(M_{\rm p}) = \frac{ (1-e^2)^\frac{3}{2} K_\star^{\ 3} P }{ 2 \pi G }
             = \frac{ M_{\rm p}^{\ 3} \sin^3 i }{ (M_\star+M_{\rm p})^2 }
\end{equation}
where $K_\star$ is the velocity amplitude of the star, $e$ is the orbital eccentricity, $P$ is the orbital period, $i$ is the orbital inclination, and $M_\star$ and $M_{\rm p}$ are the masses of the star and planet respectively. Including Kepler's Third Law and solving for the sum of the masses of the two components gives:
\begin{equation} \label{eq:mm}
(M_\star + M_{\rm p} )^2 = \frac{ 2 \pi G M_{\rm p}^{\ 3} \sin^3 i }{ (1-e^2)^\frac{3}{2} K_\star^{\ 3} P }
                         = \frac{ (2\pi)^4 a^6 }{ G^2 P^4 } \label{1c}
\end{equation}
By substituting $R_{\rm p} = a r_{\rm p}$ into the definition of surface gravity and replacing $a$ using Eq.\,\ref{eq:mm} we find that the surface gravity of the planet, $g_{\rm p}$ is given by:
\begin{equation} \label{eq:g}
g_{\rm p} = \frac{ 2 \pi }{ P } \,\frac{ (1-e^2)^\frac{1}{2} K_\star }{ r_{\rm p}^{\ 2} \sin i }
\end{equation}

Eq.\,\ref{eq:g} shows that we are able to calculate the surface gravity of a transiting extrasolar planet from the quantities $P$, $K_\star$, $e$, $i$ and $r_{\rm p}$. This can be understood intuitively because both the radial velocity motion of the star and the planet's surface gravity are manifestations of the acceleration due to the gravity of the planet. A discussion in the context of eclipsing binaries is given by \citet{Me+04mn3}.

To measure $g_{\rm p}$ using Eq.\,\ref{eq:g}, the orbital period, $P$, can be obtained from either radial velocities or light curves of the system, and is typically determined very precisely compared to the other measurable quantities. The radial velocities also give $e$ and $K_\star$, whilst the quantities $i$ and $r_{\rm p}$ can be obtained directly from the transit light curve. Note that it is also possible to constrain the orbital eccentricity from observations of the secondary eclipse of a system.


\section{Application to HD\,209458\,b}

In order to measure the surface gravity for HD\,209458\,b we need to know $r_{\rm p}$ and $i$. These quantities are standard parameters in the analysis of transit light curves. We have chosen to obtain them by modelling the high-precision HST/STIS light curve presented by \citet{Brown+01apj}. We followed Brown et al.\ by rejecting data from the first HST orbit of each observed transit.

To model the photometric data we used the {\sc jktebop} code\footnote{{\sc jktebop} is written in {\sc fortran77} and the source code is available at {\tt http://www.astro.keele.ac.uk/$\sim$jkt/}} \citep{Me++04mn2}, which is a modified version of the {\sc ebop} program \citep{PopperEtzel81aj,Etzel81conf}. \citet{Gimenez06aa} has shown that {\sc ebop} is very well suited to the analysis of the light curves of transiting extrasolar planets. Importantly for this application, {\sc jktebop} has been extended to treat limb darkening (LD) using several non-linear LD laws \citep{Me++07aa}. It also includes Monte Carlo and bootstrapping simulation algorithms for error analysis \citep{Me++04mn2,Me+04mn3}. {\sc ebop} and {\sc jktebop} model the two components of an eclipsing system using biaxial ellipsoids \citep{NelsonDavis72apj,Etzel75}, so allow for the deformation of the bodies from a spherical shape.

When modelling the data we adopted the precise orbital period of 3.52474859 days given by \citet{Knutson+07apj}. We fitted for the sum of the fractional radii, $r_\star$+$r_{\rm p}$, the ratio of the radii, $k = \frac{r_{\rm p}}{r_\star} = \frac{R_{\rm p}}{R_\star}$, the orbital inclination, and the midpoint of a transit, $T_0$. We also fitted for the LD coefficients of the star, rather than fixing them at values calculated using model atmospheres, to avoid introducing a dependence on theoretical predictions. The linear and non-linear LD coefficients are denoted by $u_\star$ and $v_\star$, respectively.

\begin{figure} \includegraphics[width=0.48\textwidth,angle=0]{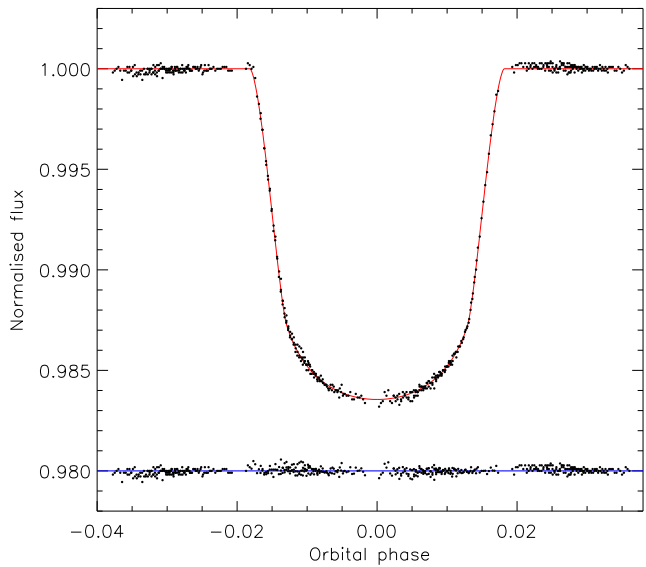} \\
\caption{\label{fig:lcfit} Best fit to the HST transit light curve of HD\,209458
using the quadratic LD law. The residuals of the fit are shown offset downwards
by 0.02 in flux for clarity.} \end{figure}

We assumed that the planet contributes no light at the optical wavelengths considered here \citep[see][]{Wittenmyer+05apj} and that the orbit is circular \citep[see][]{Laughlin+05apj,Deming+05Natur,Winn+05apj}. Given suggestions that the choice of LD law can influence the derived inclination \citep{Winn+05apj}, we obtained solutions for the linear, quadratic and square-root laws \citep{Me++07aa}. A mass ratio of 0.00056 was used \citep{Knutson+07apj} but large changes in this parameter have a negligible effect on the solution.

We have calculated robust 1\,$\sigma$ error estimates using Monte Carlo simulations (see \citealt{Press+92book}, p.684; \citealt{Me++04mn2}), which we have previously found to provide very reliable results \citep{Me++05aa,Me+05mn}. This procedure assumes that systematic errors are negligible. We find no reason to suspect that significant systematic errors remain in the HST light curve after the processing of this data described by \citet{Brown+01apj} (see Fig.\,\ref{fig:lcfit}).

The best-fitting parameters of the transit light curve are given in Table\,\ref{tab:lcresult}. The best-fitting model using the quadratic LD law is shown in Fig.\,\ref{fig:lcfit} along with the residuals of that fit. The solution using linear LD can be rejected as its reduced $\chi^2$ is substantially larger than for the other two solutions. The quadratic and square-root LD laws give very similar solutions with reduced $\chi^2$ values close to one. For our final results we adopt the mean for each parameter along with uncertainties from the Monte Carlo simulations (Table\,\ref{tab:lcresult}). These are in good agreement with the light curve solutions obtained by \citet{Gimenez06aa} and \citet{MandelAgol02apj}, both of which used the approximation that the planet is spherical.

With the orbital period given by \citet{Knutson+07apj}, the stellar velocity ampitude $K_\star = 85.1 \pm 1.0$\ms\ from \citet{Naef+04aa}, and the results of our light curve analysis (Table\,\ref{tab:lcresult}) we find the surface gravity of HD\,209458\,b to be $g_{\rm p} = 9.28 \pm 0.15$\mss. In this case the total uncertainty in $g_{\rm p}$ is due to almost equal contributions from the uncertainties in $K_\star$ and $r_{\rm p}$.


\section{Application to all known transiting extrasolar planets}

We have calculated the surface gravity values for each of the known transiting extrasolar planets (apart from HD\,209458), using data taken from the literature (Table\,\ref{tab:logg}). In several cases (marked with asterisks in Table\,\ref{tab:logg}) it was not possible to obtain $r_p$ directly from the results available in the literature. In these cases it had to be calculated from $R_p$ and $a$, resulting in an increased uncertainty. This is because $r_p$ is a parameter obtainable directly from a transit light curve, whereas additional constraints (for example using theoretical stellar models) are needed to calculate $a$ and $R_p$.

\begin{figure} \includegraphics[width=0.48\textwidth,angle=0]{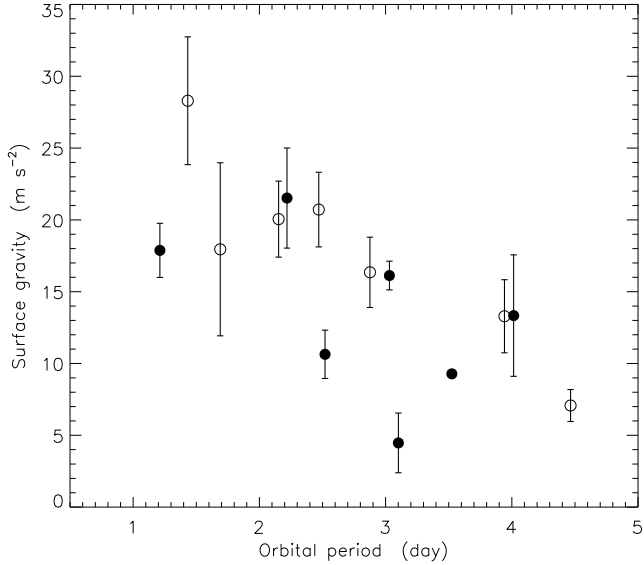} \\
\caption{\label{fig:logg} Comparison between the surface gravities and orbital
periods of the known transiting exoplanets. Filled and open circles denote the
systems in the upper and lower halves of Table\,\ref{tab:logg}, respectively.
The errorbars for HD\,209458 ($P = 3.52$\,d) are smaller than the plotted
symbol.} \end{figure}

The orbital periods and surface gravities of all fourteen transiting extrasolar planets are plotted in Fig.\,\ref{fig:logg}, and show that these quantities are correlated. The linear Pearson correlation coefficient of these data is $r = -0.70$, indicating that the correlation is significant at better than the 0.5\% level. This correlation is certainly related to that found by \citet{Mazeh++05mn} between the orbital periods and masses of the six transiting extrasolar planets then known. However, it may be that surface gravity, rather than mass or radius, is the main parameter correlated with orbital period for these objects. Theoretical calculations have shown that surface gravity is a fundamental parameter in the evaporation rates of the atmospheres of irradiated gas giant planets \citep{Lammer+03apj}.

\begin{table} \begin{center}
\caption{\label{tab:logg} Surface gravity values for the known transiting
extrasolar planets. These have been calculated using Eq.\,\ref{eq:g} with
input parameters taken from the literature.}
\begin{tabular}{l r@{\,$\pm$\,}l c c} \hline
System            & \mc{Surface gravity}              & \mc{Literature\,references} \\
                  & \multicolumn{2}{l}{m s$^{-2}$}    & $r_p$ and $i$ & $K_\star$   \\
\hline            
HD 189733         &         21.5    &    3.5          &        1      &      2      \\
HD 209458         &          9.28   &    0.15         &        3      &      4      \\
OGLE-TR-10        &          4.5    &    2.1          &        5      &      6      \\
OGLE-TR-56        &         17.9    &    1.9          &        5      &      2      \\
OGLE-TR-111       &         13.3    &    4.2          &        7      &      8      \\
TrES-1            &         16.1    &    1.0          &        9      &     10      \\
WASP-1            &         10.6    &    1.7          &       11      &     12      \\
\hline            
$^*$ HAT-P-1      &          7.1    &    1.1          &       13      &     13      \\
$^*$ XO-1         &         13.3    &    2.5          &       14      &     14      \\
$^*$ HD 149026    &         16.4    &    2.5          &       15      &     15      \\
$^*$ OGLE-TR-113  &         28.3    &    4.4          &       16      &     17      \\
$^*$ OGLE-TR-132  &         18.0    &    6.0          &       18      &     17      \\
$^*$ TrES-2       &         20.7    &    2.6          &       19      &     19      \\
$^*$ WASP-2       &         20.1    &    2.7          &       20      &     12      \\
\hline \end{tabular} \end{center}
$^*$ The surface gravity values for these objects have larger error estimates
than are needed, because their fractional radii are not available in the
literature. In these cases we have had to calculate them from $R_p$ and $a$,
which are less certain than $r_p$ because of the need to adopt an additional
constraints to calculate them (see text).
\newline {\bf References:}
(1) \citet{Winn+07aj}; (2) \citet{Bouchy+05aa}; (3) This work; (4) \citet{Naef+04aa};
(5) \citet{Pont+07aa}; (6) \citet{Konacki+05apj}; (7) \citet{Winn++07aj};
(8) \citet{Pont+04aa}; (9) \citet{Winn++07aj2}; (10) \citet{Alonso+04apj};
(11) \citet{Shporer+06xxx}; (12) \citet{Cameron+07mn}; (13) \citet{Bakos+07apj};
(14) \citet{Mccullough+06apj}; (15) \citet{Sato+05apj}; (16) \citet{Gillon+06aa};
(17) \citet{Bouchy+04aa}; (18) \citet{Gillon+07xxx}; (19) \citet{Odonovan+06apj};
(20) \citet{Charbonneau+07apj}.
\end{table}

\section{Summary and discussion}

We have shown that the surface gravity of transiting extrasolar planets can be measured from analysis of the light curve and a spectroscopic orbit of the parent star. We have analysed the HST/STIS light curve of HD\,209458 \citep{Brown+01apj} with the {\sc jktebop} code. By combining the results of the light curve analysis with published spectroscopy \citep{Naef+04aa} we find that the planet has a surface gravity of $g_{\rm p} = 9.28 \pm 0.15$\mss. We stress that this measurement does not depend on theoretical stellar evolutionary models or model atmospheres.

\begin{figure} \includegraphics[width=0.48\textwidth,angle=0]{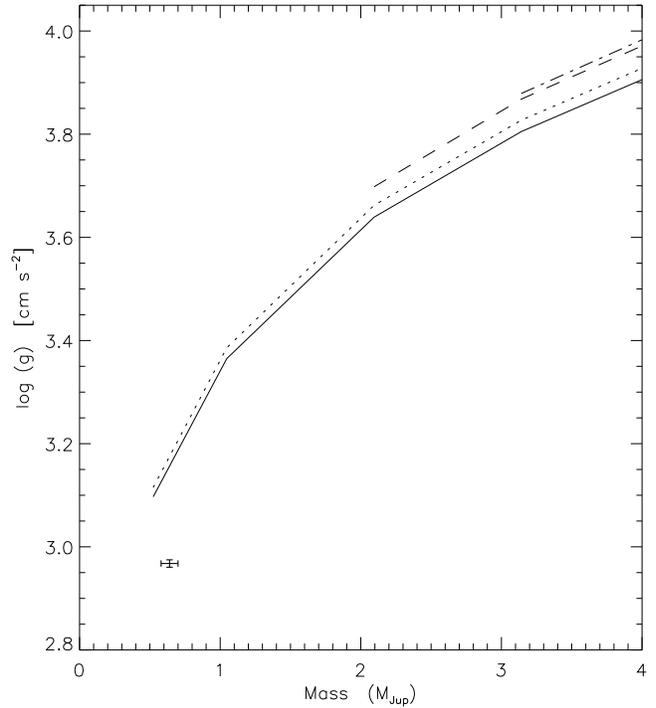} \\
\caption{\label{fig:loggmod} Plot of mass versus surface gravity for HD\,209458\,b
compared to the theoretical model predictions of \citet{Baraffe+03aa} for ages 0.5,
1.0, 5.0 and 10.0 Gyr (from lower to higher \logg).} \end{figure}

In Fig.\,\ref{fig:loggmod} we have plotted theoretical isochrones for ages of 0.5 to 10 Gyr from \citet{Baraffe+03aa} against the mass and surface gravity of HD\,209458\,b, adopting a mass of $M_{\rm p} = 0.64 \pm 0.06$\Mjup\ from \citet{Knutson+07apj}. The discrepancy between the observed and predicted surface gravity can clearly be seen. Note that these models do not include the effects of irradiation from the star.

The density of a transiting extrasolar planet is often used to compare observation with theory, but is typically
measured with a much lower precision than its surface gravity, given the same dataset. For example, the density of HD\,209458\,b derived by \citet{Knutson+07apj} is $345 \pm 50$\,kg\,m$^{-3}$. Using the mass and radius given by Knutson et al.\ leads to $g_{\rm p} = 9.1 \pm 0.9$\mss, where the uncertainty has been calculated by simple error propagation. These quantities are both much less precise and require more elaborate calculations than using Eq.\,\ref{eq:g} to find the surface gravity: $g_{\rm p} = 9.28 \pm 0.15$\mss.

We have applied Eq.\,\ref{eq:g} to each of the known transiting extrasolar planets (Table\,\ref{tab:logg}). The resulting surface gravities show a clear correlation with orbital period (Fig.\,\ref{fig:logg}) which is connected with the known correlation between orbital period and mass for these objects. We propose that surface gravity may be the underlying parameter of the correlation due to its influence on the evaporation rates of the atmospheres of short-period giant planets.

As $g_{\rm p}$ can be very precisely measured, and can be directly compared with theoretical models and used to construct model atmospheres of the planet, we propose that it is an important parameter in our understanding of short-period extrasolar giant planets. In the near future, the high-precision light curves obtained by the {\em CoRoT} and {\em Kepler} satellites will allow accurate surface gravity values to be obtained for the terretrial-mass transiting extrasolar planets which these satellites should find.

\section{Acknowledgements}

We are grateful to David Charbonneau for making the HST/STIS light curve of HD\,209458 available on his website ({\tt http://cfa-www.harvard.edu/$\sim$dcharbon/frames.html}), to Pierre Maxted for discussions, and to the referee whose report contributed significantly to improvements in this work.

JS acknowledges financial support from PPARC in the form of a postdoctoral research assistant position. The following internet-based resources were used in research for this paper: the NASA Astrophysics Data System; the SIMBAD database operated at CDS, Strasbourg, France; and the ar$\chi$iv scientific paper preprint service operated by Cornell University.


\bibliographystyle{mn_new}

\end{document}